\documentstyle[aps,prl,amsfonts,amsmath,epsf]{revtex}

\begin{document}
\twocolumn[\hsize\textwidth\columnwidth\hsize\csname@twocolumnfalse\endcsname

\epsfverbosetrue

\draft

\title{Collapsing dynamics of attractive Bose-Einstein condensates}

\author{L. Berg\'e$^1$ and J. Juul Rasmussen$^2$}

\address{
$^1$ Commissariat \`a l'Energie Atomique, CEA/DAM - Ile de France,
B.P. 12, 91680 Bruy\`eres-le-Ch\^atel, France.\\
$^2$ Ris{\o} National Laboratory, Optics and Fluids Dynamics Department, P.O.
Box 49, 4000 Roskilde, Denmark.}

\date{\today}

\maketitle

%\normalsize

\begin{abstract}
The self-similar collapse of 3D and quasi-2D atom condensates with negative
scattering length is examined. 3D condensates are shown to blow up following
the scenario of {\it weak collapse}: The inner core of the condensate diverges
with an almost zero particle number, while its tail distribution spreads out to
large distances with a constant density profile. For this case, the 3-body
recombination arrests the collapse, but it weakly dissipates the atoms. The
confining trap then reforms the condensate at later times. In contrast, 2D
condensates undergo a {\it strong collapse}: The atoms stay mainly located at
center and recombination sequentially absorbs a significant amount of particles.
\end{abstract}


\vskip1pc]

%\begin{multicols}{2}

%\narrowtext

A few years ago, Bose-Einstein condensates (BECs) were discovered
in trapped clouds of alkali atoms \cite{anderson,bradley}. Among
those, $^7$Li atoms are known to be characterized by attractive
interactions with a negative scattering length, $a_0 < 0$
\cite{bradley}. This promotes a collapse-type instability, which
yields a singular increase in the BEC wavefunction. It occurs
when the number of particles, ${\tilde N}$, exceeds a threshold value,
${\tilde N}_c \sim l_0/|a_0|$, where $l_0 = (\hbar/m\omega)^{1/2}$ is the
amplitude of zero point oscillations in the confining trap of
frequency $\omega$ and $m$ is the atom mass. BECs formed in
$^7$Li gas develop several sequences of collapse \cite{sackett}.
From the initial cooling, the condensate is first fed by the
thermal cloud of uncondensed atoms. Then, once the number of
atoms is above ${\tilde N}_c$, the condensate sharply concentrates with an
increasing density. As the density rises, inelastic
collisions as 3-body molecular recombination increase, which
arrests the collapse. The thermal cloud is next believed to
re-fill in the condensate, that reaches ${\tilde N} > {\tilde N}_c$ and then
collapses again. The cycle of collapses repeats many times until
the gas comes to equilibrium. This scenario is supported by
numerical integrations of the Gross-Pitaevskii (G.-P.) equation
\cite{kagan1,ueda}:
\begin{equation}
\label{1}
\displaystyle{i\partial_t \psi = - \Delta_r \psi + r^2 \psi - |\psi|^2 \psi -
i\eta |\psi|^m \psi + i\gamma \psi,}
\end{equation}
where $r = {\tilde r}/l_0$, $t = {\tilde t}\omega/2$, $\psi =
{\tilde \psi} (8\pi l_0^2|a_0|)^{1/2}$ are the dimensionless
time, coordinates and wavefunction of the BEC (tilde refers to
physical quantities). Here, the loss/gain mechanisms are
described by the last two terms of Eq. (\ref{1}), where $\eta \ll
1$ and $\gamma \ll 1$ are the coefficients related to the local
recombinational decrease of the condensate density and to the
flux of particles from the nonequilibrium thermal cloud to the
condensate, respectively. At leading order, we consider 3-body
recombinations with $m = 4$. The normalized number of particles,
$N = \int |\psi|^2 d{\vec r}$, is related to its physical
counterpart as $\tilde{N} = N (l_0 /8 \pi |a_0|) \simeq 86.7
N$. It was recently shown that stationary condensates defined by
$\psi_s(r,t) = \chi(r,\mu) \mbox{e}^{-i\mu t}$ with chemical
potential $\mu$ are stable whenever they satisfy $dN_{eq}/d\mu <
0$, where $N_{eq}(\mu) = \int |\chi|^2 d{\vec r}$ \cite{berge2}. Along
the curve $N_{eq}(\mu)$, 3D condensates then requires $N_{eq}
\leq N_c = 14.45$ for being stable.
 In physical units, this threshold yields the precise critical number for
stability: ${\tilde N}_c = 1252$ atoms, for
the $^7$Li parameters $a_0 = - 1.45$ nm and $l_0 = 3.16$ $\mu$m
used in \cite{bradley,sackett,ueda}.

Numerical simulations \cite{kagan1} of Eq. (\ref{1}) revealed
that, near the collapse moment, the condensate expands with a
density profile exhibiting a low-amplitude, almost flat plateau,
$|\psi|^2 r^2 \rightarrow$ const, from $r > 1$. This plateau-like
behavior was again numerically found in \cite{ueda}, where the
free collapse was described as a "black hole". Following this
scenario, the increase in the BEC density is fueled by particles
drawn from throughout the outer region ($r > 1$), while the
density outside form imploding ripples. However, it was suggested
that the asymptotic plateau in $r^2|\psi|^2$ may not be constant,
but instead diverges in time. Thus, the singular dynamics of 3D
collapsing BECs is still questionable and a self-consistent model
for BEC collapse is actually missing. Describing the structure of
such collapses is of utmost importance, in order to understand the
influence of recombination, trap confinement and re-feeding
over several collapse events.

In this letter, we examine the self-similar nature of collapsing
BECs. For 3D isotropic condensates the number of particles is
analytically shown to vanish near center and, outside, a constant
plateau in the density profile, $r^2 |\psi|^2 $, is actually
formed. From this dynamics, 3-body recombination arrests the
collapse by removing a limited number of particles from the
condensate. We also briefly investigate 2D condensates. As
justified in \cite{petrov}, quasi-2D BECs can be produced
from 3D atom clouds with a density frozen on a Gaussian shape
$\propto \mbox{e}^{-{\tilde z}^2/l_0^2}$, when the particles are
tightly confined along the longitudinal axis. In what follows, 2D and 3D
condensates are considered as isotropic, radially-symmetric
objects, since the self-compression induced by the collapse
dynamics make them have comparable sizes along each direction.

Let us first discuss the inertial regime of collapse, for which
we set $\gamma = \eta = 0$ in Eq. (\ref{1}). Collapse occurs in
the sense that the mean-square width of attractive BECs, $\langle
r^2 \rangle = N^{-1} \int r^2 |\psi|^2 d{\vec r}$, tends to zero
in finite time (see, e.g., Pitaevskii \cite{pitaevskii}). From the
inequality $N \leq \langle r^2 \rangle \int |\nabla \psi|^2
d{\vec r}$ \cite{berge1}, the vanishing of $\langle r^2 \rangle$
leads to the blow-up of the gradient norm, which in turn implies
the divergence of the integral $\int |\psi|^4 d{\vec r}$ in the
conserved Hamiltonian of Eq. (\ref{1}):
\begin{equation}
\label{2}
\displaystyle{H = \int |\nabla \psi|^2 d{\vec r} - \frac{1}{2} \int |\psi|^4
d{\vec r} + N \langle r^2 \rangle.}
\end{equation}
By virtue of the mean-value theorem $|\psi|^4 \leq \mbox{max}_r
|\psi|^2 \int |\psi|^2 d{\vec r}$, the maximum amplitude of the
wavefunction also blows up in finite time. It should be
emphasized that the blow-up generally occurs before $\langle r^2 \rangle$ reaches zero \cite{berge1}. To examine the shape of collapsing condensates, we introduce the self-similarlike substitution:
\begin{equation}
\label{3}
\displaystyle{\psi({\vec r},t) = a^{-\alpha}(t) \phi({\vec \xi},\tau)
\mbox{e}^{i\lambda \tau - i \beta \xi^2/4},}
\end{equation}
where ${\vec \xi} = {\vec r}/a(t)$, $\tau(t) \equiv \int_0^t
du/a^2(u)$ and $\beta = - a \dot{a}$ (dot means differentiation
with respect to time). Here, the parameter $\lambda \sim -\mu$
must be positive in the absence of the trap for making $\phi$
localized. The function $a(t)$ represents the BEC scale length
that vanishes as collapse develops, and $\phi({\vec \xi},\tau)$
is a regular function with amplitude of order unity. As $a(t)
\rightarrow 0$, it is assumed that $\phi$ converges to an exactly
self-similar form $\phi({\vec \xi})$, which no longer depends
explicitly on time, i.e., $\partial_{\tau} \phi \rightarrow 0$.
In this limit, the right balance between the two integrals in Eq.
(\ref{2}) requires $\alpha = 1$, in order to assure the
finiteness of $H$. The particle number $N = \int |\psi|^2 d{\vec
r}$ then reads $N = a^{D-2}(t) \int |\phi|^2 d{\vec \xi}$ and the
dynamics drastically changes following the space dimension number
$D$. Setting thus $\alpha = 1$ and plugging Eq. (\ref{3}) into
(\ref{1}) transforms the G.-P. equation into
$$
i\partial_{\tau} \phi + \Delta_{\xi} \phi + |\phi|^2 \phi +
\epsilon [\xi^2 - \xi_{T}^2]\phi - a^4 \xi^2 \phi \,+
$$
\begin{equation}
\label{4}
\displaystyle{i\eta a^{2-m} |\phi|^m \phi - i \gamma a^2 \phi = 0,}
\end{equation}
where $\Delta_{\xi} = \xi^{1-D} \partial_{\xi} \xi^{D-1}
\partial_{\xi}$, $\epsilon \equiv - \frac{1}{4} a^3 \ddot{a} =
\frac{1}{4}(\beta^2 + \beta_{\tau})$ and $\xi_{T}^2 \equiv
\epsilon^{-1}[\lambda + i\beta(D/2-1)]$ is viewed as a turning
point. After an initial stage during which the trap gathers the
particles at center, the wavefunction diverges freely on short
time scales $\Delta {\tilde t} \ll \omega^{-1}$. It becomes
hyperlocalized at $r \sim 0$ with $a(t) \rightarrow 0$, so that
the effect of trapping  can be neglected. Near the collapse
point, refeeding from the surrounding cloud is also inefficient.
Moreover, 3-body recombination does not act, as long as the BEC
radius satisfies $\eta a^{2-m} |\phi|^m \ll |\phi|^2$. In this
regime, Eq. (\ref{4}) thus reduces to the self-similarly
transformed nonlinear Schr{\"o}dinger (NLS) equation, whose
properties, accurately verified in
\cite{berge1,mclaughlin,kosmatov}, are recalled below.

In the self-similar limit $a(t) \rightarrow 0,\,\partial_{\tau} \phi
\rightarrow 0$, the time-dependent function $\epsilon$ converges to
$\beta^2/4$, which, for self-consistency, converges to a constant,
$\epsilon_0$. (ii) The solution $\phi$ in Eq. (\ref{4}) can be decomposed into
a {\it core} contribution $\phi_c$ extending in the range $\xi < \xi_T$, i.e.,
$r < r_T \equiv a(t) |\xi_{T}|$, and a {\it tail} $\phi_{T}$ defined in the
complementary spatial domain $\xi > \xi_{T}$, i.e., $r > r_T$. (iii) The
self-similar assumption $\partial_{\tau} \phi \rightarrow 0$ holds provided that
$\xi < \xi_{max} \equiv A|\xi_{T}|/a(t)$, where $A = \mbox{const} \gg 1$.
Self-similar solutions are thus limited by the boundary radius $r_{max} = A|\xi_{T}| \gg 1$. Knowing this, the solution $\phi$ decomposes as $\phi =
\phi_c|_{\xi < \xi_T} + \phi_T|_{\xi_T < \xi < \xi_{max}}$. In the long
distance domain, the nonlinearity vanishes, so that $\phi_T$ can be determined
from the linear version of Eq. (\ref{4}) by means of WKB methods \cite{berge1}.
As a result, the wavefunction $\psi$ reads near the collapse point ($\epsilon =
\beta^2/4$) as
\begin{equation}
\label{5}
\displaystyle{\psi({\vec r},t) = \frac{\mbox{e}^{i\lambda \int_0^t
\frac{du}{a^2(u)}}}{a(t)} \times}
\end{equation}
$$
\{ \phi_c(\frac{r}{a},\epsilon) \mbox{e}^{- i \beta r^2/4a^2}|_{0 \leq r < r_T}
+  \frac{C(\beta)}{(r/a)^{1+i\lambda/\beta_0}}|_{r_T < r < r_{max}}\},
$$
\begin{equation}
\label{6}
\displaystyle{|C(\beta)|^2 \simeq \frac{2\phi^2(0)}{\beta|\xi_T|^{D-2}}
\,\mbox{e}^{f(\lambda/\beta)},}
\end{equation}
$$
f(\frac{\lambda}{\beta}) = - \frac{\pi \lambda}{\beta} + (\frac{D}{2}-1)(1 +
2\ln{2}) - \frac{\lambda}{\beta} \mbox{arctan}[\frac{\beta}{\lambda}(\frac{D}{2}-1)].
$$
The scaling law $a(t)$ in the inertial range of collapse is then identified
through the continuity equation for $\psi$:
$$
\int_0^{r_{max}} \partial_t |\psi|^2 r^{D-1}dr = - 2 r^{D-1} |\psi|^2
\partial_r \mbox{arg}(\psi)|^{r_{max}}
$$
\begin{equation}
\label{7}
\displaystyle{- 2 \eta \int_0^{r_{max}} |\psi|^{m+2} r^{D-1} dr + 2 \gamma
\int_0^{r_{max}} |\psi|^2 r^{D-1}dr,}
\end{equation}
where arg$[\psi(r=0)] = 0$. By applying the solution (\ref{5}), (\ref{6}) to Eq.
(\ref{7}) and Taylor-expanding $\phi_c$ around $\epsilon = \beta^2/4 = \epsilon_0$,
the contraction scale $a(t)$ is indeed readily determined for $\eta = \gamma =
0$ from the dynamical system:
\begin{equation}
\label{9}
\displaystyle{c_1 \beta_{\tau} \simeq - \frac{c_2}{\beta}
\mbox{e}^{f(\lambda/\beta)} + D - 2,}
\end{equation}
where $c_1 \propto \mbox{Re} \int \phi_0^* \partial_{\epsilon}
\phi|_{\epsilon_0} d{\vec \xi}$ and $c_2 > 0$ are constants.

We specify the structure of the collapse for 3D and quasi-2D BECs
separately:

1 - {\it Three-dimensional condensates}: Let us first describe the
inertial range of collapse. For $D = 3$, Eq.
(\ref{9}) shows that $\beta$ rapidly attains a fixed point
$\beta_0 > 0$ corresponding to the self-similar state
$\beta_{\tau} = 0$. The scaling law $a(t) \sim (t_c - t)^{1/2}$
follows, where $t_c$ denotes the collapse moment. The
characteristics of a 3D collapse is that the number of particles
is not preserved self-similarly in the whole spatial domain,
since $N = a(t) N\{\phi\}$ with $N\{\phi\} = \int |\phi|^2 d{\vec
\xi}$. By virtue of Eq. (\ref{5}), we see that, near the collapse instant,
 $N$ behaves as $N \simeq
N_{core}(t) + N_{tail}(t) = a(t)[N\{\phi_c\}(t) + N\{\phi_T\}(t)]$, where
\begin{equation}
\label{9b}
\displaystyle{N\{\phi_c\} = 4\pi \int_0^{\xi_T} |\phi_c|^2 \xi^2 d\xi \simeq
O(1),}
\end{equation}
$$
N\{\phi_T\} = 4\pi |C(\beta)|^2 \int_{\xi_T}^{\xi_{max}} d\xi \simeq 4\pi
|C(\beta)|^2 r_{max}/a(t),
$$
so that almost all $N$ lies in the tail,
$N_{tail}(t) = 4\pi |C(\beta_0)|^2 r_{max}$, as $a(t) \rightarrow 0$. Therefore,
a 3D collapse takes place at the center of the trap where the wavefunction
$|\psi| = |\phi|/a(t)$ diverges. However, it is accompanied by an expulsion of
particles towards the large-distance domain $r \gg r_{T}(t)$. Thus, after the onset of collapse, the singularity develops, not by taking
particles from outside as in the "black hole" scenario proposed
in \cite{ueda}, but {\it by ejecting particles outward the core
domain}. This is a {\it weak collapse}, as originally defined in
Ref. \cite{zakharov} for the free NLS equation. The solution
$\psi$ blows up at center, near which the number of particles
becomes zero, i.e., $N_{core}(t) = 4\pi \int_0^{r_T(t)}
|\psi_{core}|^2 r^2dr \rightarrow 0$, as $r_{T}(t) \sim a(t)$
vanishes. Accordingly, the tail of the BEC wavefunction in the
outer domain extends in space with the time-independent density:
$r^2 |\psi|^2 \rightarrow |C(\beta_0)|^2$ = const, deduced from
Eq. (\ref{5}). This constant depends on the values of $\beta_0$
and $|\xi_T|$. For the scaling law $a(t) = a_0 (t_c - t)^{1/2}$
with $a_0$ set equal to the unity without loss of generality,
$\beta_0 = 1/2$ and numerical integrations of the 3D NLS equation
with no trap \cite{kosmatov} emphasize the values $\phi_0 \equiv
\phi(0) = 1.39$ and $\lambda = 0.545$, which yield $|\xi_T| \sim
3.1$ and $|C(\beta_0)|^2 \simeq 0.17 \sim 0.2$. A tail
contribution $r^2 |\psi|^2/|\psi(t=0)|^2$ about 0.2 in magnitude
for $r \geq 1$ seems compatible with the density profiles
computed in Refs. \cite{kagan1,ueda} in the vicinity of the
collapse moment. The formation of stationary plateau-like density
profiles was numerically confirmed in \cite{mclaughlin,kosmatov}
by means of very fine numerical schemes that solved the rescaled
equation (\ref{4}) with high accuracy and could access huge
growths in $|\psi|^2$. In contrast, the same plateau was claimed
to diverge in time by only refining the spatial grid in
\cite{ueda}. We suspect that numerical computations in
\cite{ueda} suffered serious lack of resolution, which prevented
the authors from concluding correctly on the constancy of
$r^2 |\psi|^2$ at large distances.

We now discuss the dissipative/gain regime, that involves collisional losses 
and feeding by the thermal cloud. The influence of these two effects
on the condensate particle number is described by the last two
terms in the continuity equation (\ref{7}). By introducing the
self-similar shape (\ref{5}), these terms are found to read $-2
\eta a^{1-m} \int_0^{\xi_T} |\phi_c|^{m+2} \xi^2 d\xi$ and
$2\gamma a \int_{\xi_T}^{\xi_{max}} |\phi_T|^2 \xi^2 d\xi$ for $a(t) \ll 1$,
respectively. Hence, 3-body recombination mainly acts on the core
part of the solution $|\psi_c| = |\phi_c|/a(t)$, whereas
re-feeding by the surrounding cloud is efficient in the outer
region, where the main amount of particles is residing, with $2\gamma \int_0^{r_{max}} |\psi|^2 r^2dr \simeq 2 \gamma
N_{tail}(t) $.

Collisions begin to be active when $a(t)$ decreases so much
that $\eta a^{2-m} |\phi|^m \simeq |\phi|^2$ [Eq. (\ref{4})],
i.e., this contribution saturates the blow-up induced by the cubic
nonlinearity, for $m > 2$. By means of this relation, the number of particles $\Delta
N_{loss}$ lost from the condensate during one collapse event is
then yielded by that in the core region,
$N_{core}$. Inelastic collisions are thus able to stop the
collapse, but they cannot remove a lot of
particles, the major part of $N \simeq N_{tail}$ starting to be
transferred to large distances via the weak collapse. Explicitly,
we find by using Eq. (\ref{5}) expressed with arbitrary
$m$ and $D$:
\begin{equation}
\label{10}
\displaystyle{\Delta N_{loss} \simeq - 2^{D-1} (\phi_0
|\xi_{T}|)^D \pi \eta^{(D-2)/(m-2)}/D.}
\end{equation}
For $D = 3$ and $m = 4$, the particle loss is estimated as $\Delta
N_{loss} \simeq 4\pi \sqrt{\eta} \phi_0^3 \xi_T^3 /3 = 0.5 $, when
we choose the value $\eta = 2.2 \times 10^{-6}$ physically
justified in \cite{ueda} and employ $\phi_0 = 1.39$ and
$\xi_{T} = 3.1$. Thus, in physical units, the number of
atoms removed during a single blow-up event is $\Delta {\tilde
N}_{loss} = 44$ atoms. This loss of particles agrees with recent
numerical observations \cite{saito}. Here, approximately 300
particles were lost from a condensate of initial number $N_0 = 1260$ atoms, in a sequence of 5-6 individual blow-up events with peak amplitudes $|\psi|_{max} \sim \eta^{1/2-m}$. As discovered by Vlasov {\it et al.} \cite{vlasov} and confirmed in \cite{kosmatov}, an untrapped, 3D collapsing field is damped within
a very short time window, in which the field
experiences secondary blow-up events of duration $\sim
\eta^{2/m-2}$, as long as nonlinear dissipation remains active. These 
arise while most of the atoms form a broad disribution surrounding the collapsing core and are directed outwards.
Collapse is not arrested abruptly, but it is distributed through a series of individual spikes. This "distributed
collapse" \cite{vlasov} resembles the "intermittent implosions"
numerically detected in \cite{saito}. Note that these intermittent implosions,
which overlap as one collapse sequence over a trap period, may
not be resolved in real experiments. Their integrated effect should only
yield a smooth decrease in ${\tilde N}$.

Once collapse is limited, it is necessary to understand the
mechanism underlying reformation and periodic resurgences of
spiky amplitudes in the condensate \cite{sackett}. We emphasize that after the collapse stage most
particles lie in the outer domain $1 \leq r < r_{max}$, where the
atoms feel both the trap curvature and re-feeding by
the thermal cloud. The latter re-injects particles into the
BEC according to  $N(t) \simeq N_{tail} \mbox{e}^{2\gamma
t}$, i.e., on time scales $t \simeq (2\gamma)^{-1} \ln{(1+\Delta N/N_0)}$.
For reasonable values of $\gamma = 3 \times 10^{-3}$ \cite{ueda},
the physical time for re-filling the condensate with $\Delta N/N_0 \approx 0.25$
is too long for justifying the spiky oscillations occurring periodically
at times $\sim \omega^{-1}$, mentioned in, e.g., \cite{saito}. We have
thus to analyze the stage of trap reconfinement at earlier instants for
which refeeding is negligible. To this aim, we define the centroid of
the expanded structure by ${\vec X}(t) \equiv N^{-1} \int {\vec r} |\psi|^2
d{\vec r}$. From algebraic
manipulations of Eq. (\ref{1}) with $\gamma = \eta = 0$, this centroid
is found to obey the relation $\ddot{\vec X} + 4{\vec X} = {\vec 0}$.
Simple phase transformation \cite{berge3} keeps equation (\ref{1}) unchanged when passing over to the frame
moving with $X(t)$.
It is then easy to conclude that, from a reference coordinate
${\vec X(t_0)} \neq {\vec 0}$ that locates the BEC expanded in
the outer domain at a given time $t_0$, the condensate will return
to the center of the trap over physical times
$= \pi/2\omega \sim \omega^{-1}$ after the collapse moment.
Once reconfined with $N < N_c$, the condensate is close to
an equilibrium, that is described by the ground state solution
$\psi_s = \chi(r,\mu) \mbox{e}^{-i\mu t}$ of the G.-P. equation.
As shown in Ref.\cite{berge2} (see also \cite{dewel}),
ground states are stable with $dN_{eq}/d\mu < 0$ for rather broad condensates ($0.72 < \mu < 3$). They are unstable and collapse
with $dN_{eq}/d\mu > 0$ for rather narrow condensates
($\mu < 0.72$). In this range, the threshold
$N_c > N_{eq}$ decreases with $|\mu|$ as $N_c(\mu) \leq 18.94/|\mu|$. It is thus highy problable
that under the strong compression induced by the trap, the equilibrium state confined at center
corresponds to a narrow, unstable $\psi_s$ undergoing
a new collapse sequence with a lower critical number of
particles. This may repeat over several trap periods,
until the refeeding becomes a key-player for making
the resulting BEC bifurcate to a stable equilibrium
reached at a weak number of atoms \cite{sackett}.

2- {\it Quasi-two-dimensional condensates}: The dynamics
significantly changes when one considers quasi-2D BECs.
For $\eta = \gamma = 0$, the scaling law $a(t)$ inferred from Eq.
(\ref{9}) with $D = 2$ indeed behaves with a twice-logarithmic
correction: $a(t) \simeq a_0
\sqrt{t_c-t}/\sqrt{\ln{\ln{[1/(t_c-t)]}}}$
\cite{berge1,mclaughlin,kosmatov}. Since $\beta \simeq
\pi/\ln{\tau(t)} \rightarrow 0$ as $\tau(t) \simeq \ln{[1/t_c-t]}
\rightarrow +\infty$, the exponential contribution in the tail
amplitude of Eq. (\ref{6}) decreases to zero, while the core
boundary $\xi_T = \sqrt{\lambda/\epsilon}$ increases slowly to
infinity when $t \rightarrow t_c$. As a result, collapse takes
place with a core solution $\phi_c$ providing the principal
contribution in the wavefunction $\psi$. $N$ is mainly given by
$\int |\phi_c|^2 d{\vec \xi}$, which relaxes to the critical
value $N_c = \int |\chi|^2 d{\vec \xi} \simeq 11.7$.
Below this number, stationary condensates are stable
\cite{berge2}. So, most of particles stay located around the
center of the trap, when collapse develops. This scenario meets
the definition of the so-called {\it strong collapse}, which
contrasts with the 3D weak collapse that promotes a leak of
atoms to large distances. The relative variations in the
particle number due to collisional losses are governed by the
estimate $\Delta N_{loss} \simeq -2 \phi_0^{-2} \eta a^{2-m} \int
|\phi_c|^{m+2} d{\vec \xi}$. As $a(t)$ decreases, the part burnt by
recombination is here estimated by $\Delta N_{loss}/N
\simeq - 2/(1 + m/2)$, when we apply the evaluation $\eta a^{2-m}
\approx |\phi|^{2-m}$ and model $\phi$ around a Gaussian with
intensity $I_0$: $\phi = \sqrt{I_0} \mbox{e}^{-\xi^2/2}$. For $m
= 4$, 3-body recombination then removes about more than one-half
of the condensed atoms. Although rough, this estimate does not
explicitly vary with the coefficient $\eta$, as already indicated by Eq. (\ref{10}), and it slightly decreases
with $m$, in agreement with the numerical observations of Ref.
\cite{kosmatov}. Collapse is thus arrested by 3-body
recombinations that sequentially burn a substantial amount of
atoms per collapse event. This agrees with the numerical
results of Refs. \cite{kosmatov,vlasov}, from which $\Delta
N_{loss}$ is about $0.25-0.3 \times N_c$.

In conclusion, we have shown that BECs with attractive
interactions collapse with a mean radius contracting like
$\sqrt{t_c - t}$ at leading order, where $t_c$ denotes the
collapse moment. In 3D, {\it the collapse is weak}: the amplitude
of the wavefunction blows up at center, while particles are expelled outwards with a constant density
profile $r^2 |\psi|^2 \rightarrow$ const. Recombination losses
limit the collapse, but they only damp a few percents of the
atoms at each blow-up. Several blow-up events develop
within one collapse sequence. The condensate is then reconfined by the trap 
and can further undergo more collapse cycles, as long as a stable state has not been reached.
Quasi-2D condensates are subject to {\it a strong collapse}, in
the sense that the number of particles remains mostly confined at
center. In this case, recombination removes a significant
fraction (up to 0.5) of particles per collapse event.\\
This work was partly supported by the
Danish Natural Sciences Foundation (SNF-grant 9903273).

%\end{multicols}


\vskip -0.5cm

\begin{references}

\bibitem{anderson}
M.H. Anderson {\it et al.}, {\em
Science} {\bf 269}, 198 (1995); K.B. Davis {\it et al.}, {\em Phys. Rev. Lett.}
{\bf 75}, 3969 (1995).

\bibitem{bradley}
C.C. Bradley, C.A. Sackett, J.J. Tollett, and R.G. Hulet, {\em Phys. Rev.
Lett.} {\bf 75}, 1687 (1995). C.C. Bradley, C.A. Sackett, and R.G. Hulet, {\em
Phys. Rev. Lett.} {\bf 78}, 985 (1997).

\bibitem{sackett}
C.A. Sackett, J.M. Gerton, M. Welling, and R.G. Hulet, {\em Phys. Rev. Lett.}
{\bf 82}, 876 (1999).

\bibitem{kagan1}
Yu. Kagan, A.E. Muryshev, and G.V. Shlyapnikov, {\em Phys. Rev. Lett.}
{\bf 81}, 933 (1998).

\bibitem{ueda}
A. Eleftheriou and K. Huang, {\em Phys. Rev. A} {\bf 61}, 043601 (2000); See also M. Ueda and K. Huang, {\em Phys. Rev. A} {\bf 60}, 3317 (1999).

\bibitem{berge2}
L. Berg\'e, T.J. Alexander, and Yu.S. Kivshar, {\em Phys. Rev. A}
{\bf 62}, 023607 (2000).

\bibitem{petrov}
H. Gauck {\it et al.}, {\em Phys. Rev. Lett.} {\bf 81}, 5298
(1998); A.I. Safonov {\it et al.}, {\em Phys. Rev. Lett.} {\bf
81}, 4545 (1998). See also D.S. Petrov, M. Holzmann, and G.V.
Shlyapnikov, {\em Phys. Rev. Lett.} {\bf 84}, 2551 (2000) and
references therein.

\bibitem{pitaevskii}
L.P. Pitaevskii, {\em Phys. Lett. A} {\bf 221}, 14 (1996).

\bibitem{berge1}
L. Berg\'e, {\em Phys. Rep.} {\bf 303} 259 (1998); J. Juul Rasmussen and K.
Rypdal, {\em Phys. Scr.} {\bf 33}, 481 (1986).

\bibitem{mclaughlin}
D.W. McLaughlin {\it et al.}, {\em Phys. Rev.
A} {\bf 34}, 1200 (1986); B.J. LeMesurier {\it et al.}, {\em Physica D} {\bf 31}, 78 (1988); {\bf 32}, 210 (1988).

\bibitem{kosmatov}
N.E. Kosmatov, V.F. Shvets, and V.E. Zakharov, {\em Physica D} {\bf 52}, 16
(1991).

\bibitem{zakharov}
V.E. Zakharov and E.A. Kuznetsov, {\em Sov. Phys. JETP} {\bf 64}, 773 (1986).

\bibitem{saito} H. Saito and M. Ueda, arXiv:cond-mat/0002393 (2000).

\bibitem{vlasov} S.N. Vlasov, L.V. Piskunova, and V.I. Talanov, {\em Sov. Phys. JETP} {\bf 68}, 1125 (1989).

\bibitem{berge3}
L. Berg\'e, {\em Phys. Plasmas} {\bf 4}, 1227 (1997).

\bibitem{dewel} C. Huepe, S. M\'etens, G. Dewel, P. Borckmans, and M.E. Brachet, {\em Phys. Rev. Lett.} {\bf 82}, 1616 (1999).

\end{references}
\end{document}